\documentclass[twocolumn,english,superscriptaddress]{revtex4-1}
\usepackage[T1]{fontenc}
\usepackage[latin9]{inputenc}
\usepackage{color}
\usepackage{amsmath}
\usepackage{babel}
\usepackage{graphicx}
\begin{document}

\title{Spin waves in periodic antidot waveguide of complex base }

\author{Santanu Pan}
 \affiliation{Department of Condensed Matter Physics and Material Sciences, S. N. Bose
National Centre for Basic Sciences, Block JD, Sector III, Salt Lake,
Kolkata 700 106, India}

\author{Jaros\l aw W. K\l os }
\email{klos@amu.edu.pl}
\affiliation{Faculty of Physics, Adam Mickiewicz University in Poznan, Umultowska
85, 61-614 Pozna\'{n}, Poland }

\author{Szymon Mieszczak}
\affiliation{Faculty of Physics, Adam Mickiewicz University in Poznan, Umultowska
85, 61-614 Pozna\'{n}, Poland }

\author{Anjan Barman}
\affiliation{Department of Condensed Matter Physics and Material Sciences, S. N. Bose
National Centre for Basic Sciences, Block JD, Sector III, Salt Lake,
Kolkata 700 106, India}

\author{Maciej Krawczyk}
\affiliation{Faculty of Physics, Adam Mickiewicz University in Poznan, Umultowska
85, 61-614 Pozna\'{n}, Poland }

\begin{abstract}
We consider the planar magnonic waveguide with a periodic sequence
of antidots forming zig-zag pattern, where two neighboring antidots
are shifted towards the opposite edges of the waveguide. This system
has a complex base with two antidots in one unit cell. The Brillouin
zone is here two-times narrower than the Brillouin zone for the waveguide
without displacement of antidots. We have shown that for dispersion
relation folded into narrower Brillouin zone, new frequency gap can
be opened and their width can be controlled by the shift of the antidots.
We found that, the different strength of spin wave pinning at the
edges of the periodic waveguide (and their antidots) determines the
dependence of the width of gap on the shift of antidots. For the systems
with completely free or ideally pinned magnetization, these dependencies
are qualitatively different. We have found an optimum shift of antidot for maximzing the
width of the gap for the system with pinned magnetization. More interestingly, we notice that
for this kind of geometry of the structure, majority of the modes
are doubly degenerate at the edge of Brillouin zone and have a finite
group velocity at the very close vicinity of the edge of Brillouin
zone, for larger values of antidot shift. This empowers us to design
magnonic waveguide to steer the spin waves.
\end{abstract}
\maketitle

\section{Introduction }

The waveguides and transmission lines are important components of
radio-frequency \cite{Yeo10}, photonic/optical \cite{Cap13,Min06,Hun09} and
magnonic \cite{Khi12,Wah11,Sad15} integrated systems for data communication
and processing. The most obvious role of this element is transmission
of signals between different parts of the system. However, the waveguides,
in which the coherent waves propagate, are more sophisticated elements.
The geometry of the waveguide determines the quantization of the modes
(confined in the cross-section of the waveguide) and their dispersion
relation (dependence of the eigenfrequency on wave vector)describes
the propagative properties. The adjustment of structural parameters
of magnonic systems or the application of external bias field allows to
tailor and continuously control the properties of the waveguides important
for their dynamical characteristics. One can change the number of
modes for given frequency (determined by the position of branches
of dispersion relation corresponding successive modes) and control
their group delay \cite{Kob02,Sch13} (given by the group velocity
-- resulting from the slope of dispersion relation). The molding and
controlling of phase delay \cite{Her04,Au12} is the main working
principle for wave-based logic systems occurred due to the wave interference
in the networks of waveguides \cite{Li08}. In these systems, the
difference of phases of the waves at the junction of two waveguides
determines the conditions for constructive or destructive interference
which corresponds to high or low level of the output signal. The more
sophisticated processing of magnonics signals can be achieved in the
waveguides with continuously changing width \cite{Vla09}, in arrays
of coupled waveguides \cite{Sad15} or in the waveguides with dynamically
applied periodic magnetic field \cite{Chu10}. 

In magnonics \cite{Kra14,Chu15,Kim10}, the dynamics of SWs
in partially confined geometries (such as waveguides) is much more
complicated phenomena than the dynamics of the excitation of different
kinds of waves (e.g. electromagnetic waves or  elastic waves).
It results from the concurrence of two kinds of interactions (dipolar
and exchange one). Due to the presence of dipolar interaction, the
anisotropic \cite{Dav15b} and nonreciprocal \cite{Di13,Mru14} SW propagation is quite easily achievable in microstructured magnonic
system. The anisotropy of spin propagation controlled by magnetic
field can be used to design the magnonic multiplexers \cite{Dav15a,Vog14}.
The other advantage of SWs, as a carrier of information, is
the smaller size of magnonic structures and devices comparing to their
other counterparts operating on the electromagnetic waves of the same
frequency. For future applications, particular attention should be
paid on magnonic nanostructured systems which are based on the exchange
SW of wave length of few nanometers and the frequencies in
 few tens of GHz range. In this regime, the dipolar
interaction is dominated by exchange interaction and, in general,
the most of the features of magnonic systems is the same as in rescaled
photonic counterparts. However, there are still some fundamental differences
between these two media. Some of the most important differences can
be listed as follows: (i) SW propagations is limited to the
magnetic material only, (ii) boundary conditions for exchange SWs at the interfaces with nonmagnetic media are determined by surface
magnetocrystalline anisotropy resulting from the physical and chemical
states of the surface. Therefore, the surface anisotropy is an additional
factor, to the geometry and to the bulk material parameters, of controlling the
SW spectrum in waveguides at the nanoscale.

The waveguides can have different forms. However, in integrated systems,
fabricated by top down lithographic techniques, the planar structures
are the most common solutions. Therefore, a great interest is focused on planar magnonic waveguides where in-plane dimensions are much
larger than the thickness and in low frequency range so that we can
neglect the out-of-plane quantization of SWs. One of the effective
ways of tailoring dispersion relation is a periodic modulation of
structural or material parameters of the system. The periodicity in
magnonic waveguides can be introduced in various ways \cite{Lee09,Tka12,Obr13,Bai11}.
The simplest method, used for pattering of planar waveguides, is to
introduce a periodic sequence of antidots \cite{Klo14}. For very
long waves this procedure can be understood just as a molding of effective
material parameters but for the shorter wavelengths, comparable to
the period of the structure, the Bragg reflection on periodic pattern
results in folding of dispersion relation to the first Brillouin zone
(BZ). This effect can lead to the opening of frequency gaps in the
SW spectrum of the waveguide. The frequency gaps can be observed
easily in lower frequency range where only one mode exists in the whole
range of wave number. The folding of dispersion branches results in
their crossings and anti-crossings inside BZ. The anti-crossing of
modes gives the possibility to open frequency gaps for higher frequencies
where many modes may exist. The presence of frequency gaps at desired
position having appropriate width is the working principle of spectral filters.

In this study, we will investigate planar magnonic waveguides periodically
patterned by the sequence of antidots. Apart from antidot
magnonic waveguide with a single ferromagnetic material, waveguides
consisting bi-component material, i.e., antidots filled with some
other ferromagnetic material with contrasting magnetic parameters
also shows a promising band structure with tunable band gap. However,
the advantages of magnonic crystal with air holes surrounded by magnetic
material over bi-component one can be noted as: (i) larger contrast in magnetic parameters, e.g., saturation magnetization
($M_{s}$) and exchange length ($l_{ex}$) helps to obtain band structure
with distinctive gaps in the spectrum, (ii) comparatively easier fabrication
process, and (iii) cost effective method as only single magnetic material
is required. 

In dipolar interaction regime, shape of the antidot plays a consequential
role because of significant variation in the demagnetizing field in
 dot and antidot structures. In our case, due to the nanoscale
magnonic structures with few nm length scale, exchange interaction
dominates over dipolar interaction in the frequency regime of few
tens of GHz. For higher SWs frequencies, the shape
of the antidot is not very important in reshaping the magnonic band
spectrum if the cross-sectional area of antidot remains constant as
reported earlier \cite{Klo14}. However, on the other hand, this type
of system is very sensitive to broken symmetry (shift of the antidot
with respect to the centre of the waveguide) \cite{Klo14}. Therefore,
we are interested in investigating the intriguing role of broken mirror
symmetry in this kind of system.

The considered periodic system has a complex base containing two antidots per one period. The antidots
of same sizes are distinguished (in one unit cell) by opposite shifts
with respect to the long-axis of the waveguide. This procedure allows
to introduce structural changes gradually and to transit the system
from the case where the complex base is artificial (we have two indistinguishable
elements in the base and the assumed periodicity is artificially doubled)
to the case where the spatial separation between neighboring antidots
is vital. We investigate the impact of this kind of structural changes
on the SW spectrum. We paid particular attention on the tunability
of the width of the magnonic band gaps and tailoring the group velocity.
We check the role of magnetization pinning/depinning \cite{Klo12}
(on the edges of the waveguide and  antidots) on the mentioned dynamical properties.We
performed numerical studies of considered structures using two different
numerical techniques: plane wave method (PWM) and micromagnetic simulation
(MS) to crosscheck the outcomes and to obtain the whole set of complementary
results \textendash{} each of those methods has specific advantages
and limitations.

The manuscript is organized in the following way. In the next section
\textquoteleft Structure and Model\textquoteright , we describe the
considered structures of antidot waveguides and discuss the physical
models we solved using the numerical calculations. The section \textquoteleft Results
and Discussion\textquoteright{} contains the outcomes of numerical
studies complemented with detailed discussion. In the last section
\textquoteright Conclusion\textquoteright{} we summarize our results.

\section{Structure and model}

We study planar quasi one-dimensional (1D) magnonic waveguide
with a series of antidots disposed periodically in zig-zag like manner, as shown in Fig.
1. It possesses the form of an infinitely long 1D stripe with thickness
equals to $1\text{ nm}$. Square
antidots with sides $s$ of $6\text{ nm}$ are placed asymmetrically
along the waveguide. We keep the width of the waveguide  fixed
at $45\text{ nm}$.  The distance between the centers of antidots, measured along the waveguide $a$ was set to $15\text{ nm}$. The alternative shifts of the antidots  towards the edges of waveguide $w$ transform the system into the periodic waveguide of complex base with two identical antidot
in one unit cell (see Fig. 1).  We varied the
value of $w$ from $0\text{ nm}$ (complex base is artificial and waveguide
possesses exact mirror symmetry with respect to the long-axis of the
waveguide) to 18 nm (complex base with two identical elements and
broken mirror symmetry) with a regular increment.
 
The waveguide here
is made of Permalloy ($\text{Ni}_{80}\text{Fe}{}_{20}$)
with a saturation magnetization $M_{s}=0.8\times10^{6}\text{ A/m}$,
and an exchange length $l_{ex}=5.69\text{ nm}$. The value of gyromagnetic
ratio $\gamma=175.9\text{ GHz/T}$ was assumed in the calculations.
A bias magnetic field of $\mu_{0}H_{0}=1 \text{ T}$ is applied along the
$x$ direction to saturate the sample along the stripe length. It
is strong enough to fully saturate the magnetization and make collinear
arrangement of spins near edges of the waveguide. For the geometry considered here,
the SWs propagate with the wave vector $\mathbf{k}$ parallel to direction of external field $\mathbf{H}_0$ (so called backward volume geometry). The uniform SW amplitude along the thickness is assumed due to large  ratio between thickness
and lateral dimensions. 

\begin{figure}[!ht]
\includegraphics[width=8.5cm]{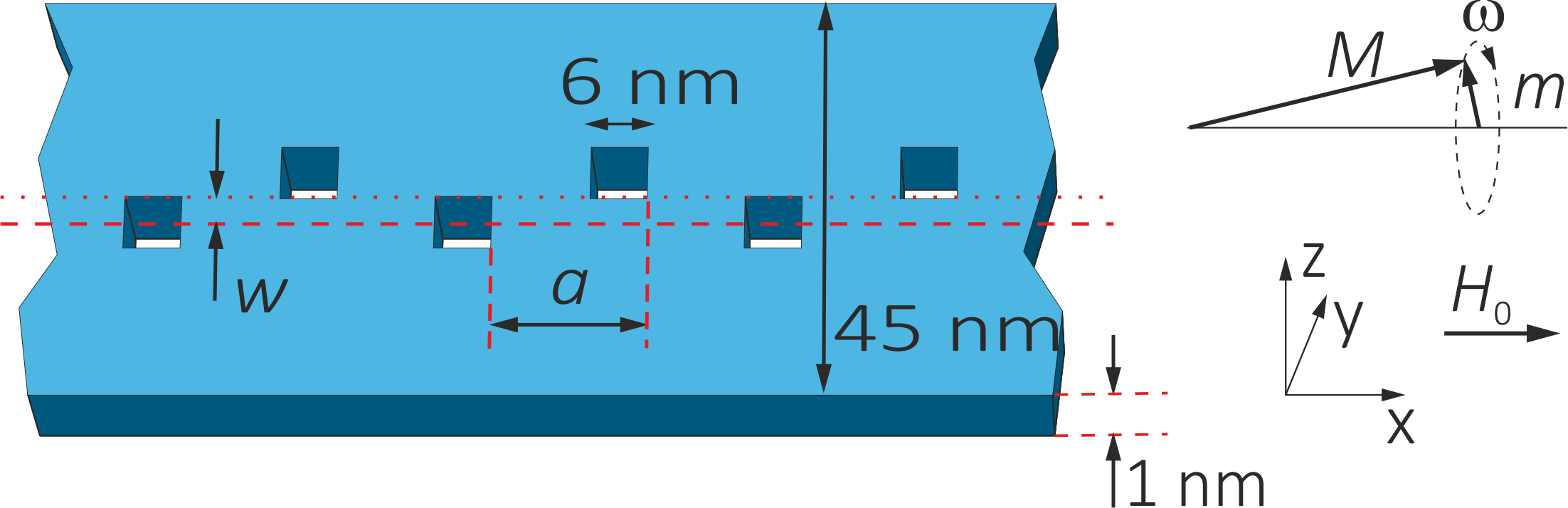}
\caption{\label{fig:Structure-of-periodically}Structure of periodically patterned
$\text{Ni}_{80}\text{Fe}_{20}$ (Py) nanowire with two square antidots
in one unit cell. The antidots are alternatively shifted from the
long axis of the waveguide (marked by dotted line) towards its edges
by the distance $w$. The period of the structure is a doubled separation
between neighboring antidots $a$, measured along the waveguide. The
external magnetic field $\mu_{0}H_{0}=1\text{ T}$ is applied along the waveguide. }
\end{figure}

We performed two different types of numerical calculations namely,
finite difference method based MS using
OOMMF \cite{Don99}, and PWM using home-built
FORTRAN code. Both these methods solve the Landau-Lifshitz-Gilbert
equation:
\begin{eqnarray}
\frac{d \mathbf{M}\left(\mathbf{r},t\right)}{d t}=\gamma\mu_{0}\mathbf{M}\left(\mathbf{r},t\right)\times\mathbf{H}_{\rm{ef\!f}}\left(\mathbf{r},t\right)+\nonumber\\
-\frac{\alpha}{M_{s}(\mathbf{r})}\mathbf{M}\left(\mathbf{r},t\right)\times\frac{d\mathbf{M}\left(\mathbf{r},t\right)}{dt}
\end{eqnarray}
where $\mathbf{r}$ and $t$ are position vector and time, respectively.
The symbols $\mu_{0}$ and $\gamma$ denote the free space permeability and
gyromagnetic ratio, respectively. There
are two torque terms present on the right-hand side of the equation.
The first term corresponds to the torque inducing the precessional
dynamics of magnetization vector $\mathbf{M}$ and the second one is responsible for damping process ($\alpha$
being the Gilbert damping coefficient). Value of $\alpha$ is neglected
in PWM calculations while a very small value of $\alpha=0.0001$
is assumed in MS allowing magnetization precession for a long time.
The field $\mathbf{H}_{\rm{eff}}$ is the total effective magnetic field which consists
of external bias magnetic field $\mathbf{H_{0}}$, exchange field
$\mathbf{H}_{\rm{ex}}=\nabla l_{\rm{ex}}^{2}(\mathbf{r})\nabla\mathbf{M}$, and demagnetizing
field $\mathbf{H}_{\rm{dem}}$. Magnetization as a function of real space
and time i.e. $\mathbf{M}\left(\mathbf{r},t\right)$ and as a function
of reciprocal space and frequency, i.e. $\mathbf{M}\left(\mathbf{k},f\right)$
are obtained from MS and PWM, respectively. Using MATLAB subroutine
program \cite{Kum12}, we analyzed the data obtained from OOMMF to
get $\mathbf{M}\left(\mathbf{k},f\right)$. The postprocessing method
is described elsewhere \cite{Klo14}. 

Magnetization pinning is observed to play an important role in opening
magnonic band gaps in this type of antidot waveguides \cite{Klo12}.
In general, intrinsic dipolar pinning due to demagnetizing field at
material/air interface affects the SW spectrum in the dipolar
regime. On the other hand, the state (pinning) of the Py/air boundary
depends on the fabrication process extending over few nm length scale
regimes. This type of pinning does not impact much in case of dipolar
interaction but critically affects the exchange SWs. Therefore,
we assume magnetization pinning of various strengths at Py/air interface
in the calculation process. Pinning in OOMMF is introduced by freezing
the magnetization direction (along $x$ direction) over a finite thin
area around Py/air interface. We create a mesh in the OOMMF calculation
using cell size $1.5\times1.5\times1\text{ nm}^{3}$ along $x$, $y$
and $z$ direction, respectively. 1D periodic boundary condition was
applied along the stripe length and total
simulation time was kept at $4\text{ ns}$ in MS for higher frequency
resolution. Pinning in PWM process is intrinsic and applied exactly
at the interface. In case of MS, the pinning is applied in the edge cell (of finite size). This may affect the results to a small
extent. However, both the methods give  similar output results
as demonstrated in previous studies \cite{Klo14,Klo12}.

\section{Results and discussion}

\begin{figure}[!ht]
\includegraphics[width=8.5cm]{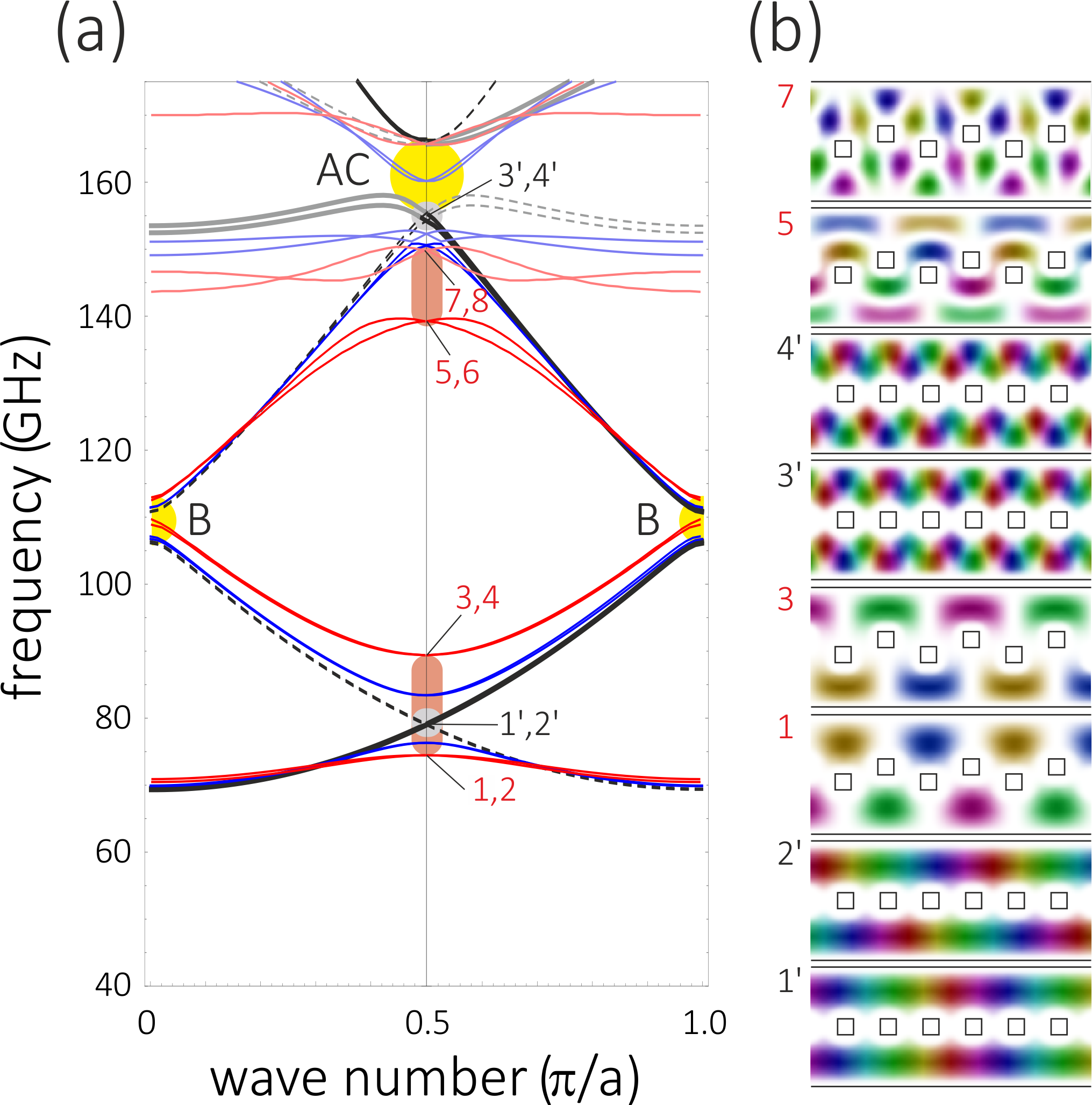}
\caption{(a) The dispersion relation for the waveguide presented in Fig. 1 for
different values of displacement of antidots: $w=0,\text{ }1.5,\text{ }3.0~\text{nm}$
\textendash{} black (gray), blue (light blue) and red (light red)
lines, respectively, calulated with plane wave method. The darker (lighter) colors of lines refer to
the modes originating from first (second) mode of uniform waveguide.
The black and gray solid lines denote the dispersion of the waveguide
with the antidots placed inline ($w$ = 0) where the unit cell is twice
 smaller (period is equal to $a$). By selecting the period equals
to $2a$ we obtain the artificial folding of dispersion relation (black
and gray dashed lines). Crossing in this folded dispersion, after
introducing the displacement of antidots ($w>0$), leads to appearance of
new gaps (pink bars at $k=\pi/2a$). These gaps are substantially
different from the Bragg gaps (label B) already opened at the center
of Brillouin zone or the gaps resulting from anti-crossing (label
AC) of the modes originating from first and second mode of homogeneous
wire of half-width. (b) Profiles of the out-of-plane components of
selected spin wave modes (for $w=0,\text{ }3.0\text{ nm}$) marked
in (1). The sort of color and color saturation correspond to phase
and amplitude of spin waves, respectively.}
\end{figure}

The periodic displacement of antidots illustrated in Fig. 1 doubles
the periodicity of the waveguide to $2a$ in reference to the system
with antidots placed inline ($w=0$). This doubling of period results
in the folding of dispersion relation to one half of its initial width in reciprocal space. In this narrowed BZ, the new magnonic band gaps appear.
We trace the opening and gradual widening of these gaps with increasing displacement $w$ for the system with pinning. In Fig. 2a we plotted
the SW dispersion for few values of the displacement $w$ calculated with the aid of PWM.

\begin{figure*}[!ht]
\includegraphics[width=1\textwidth]{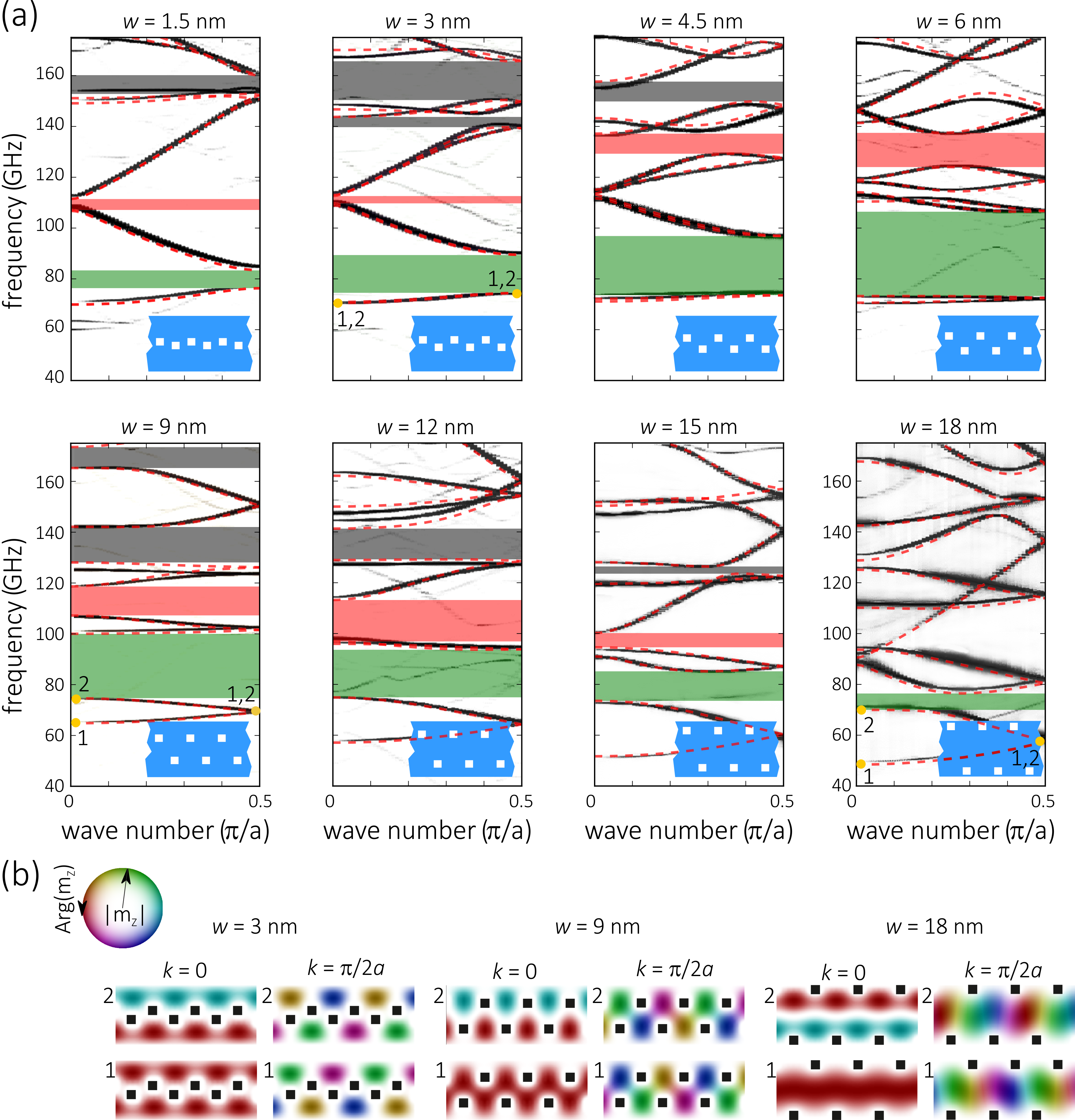}
\caption{(a) Spin wave dispersion relation for different displacements of antidots
$w=1.5,\text{ }3,\text{ }4.5,\text{ }6,\text{ }9,\text{ }12,\text{ }15,\text{ }18\text{ nm}$
calculated with the aid of micromagnetic simulation  and
plane wave method (dashed lines). We assumed the pinning of magnetization
at the edges of the waveguide and the antidots. The green and pink
bars denote the lowest magnonic gaps resulting from the presence of
the complex unit cell. (b) Spatial profiles of two lowest spin wave
modes presenting the out-of-plane component of magnetization at the
center and at the edge of first Brillouin zone.}
\end{figure*}

The case $w=0$ corresponds to the waveguide with all antidots placed
inline and the period equal to $a$. In the considered (Fig. 2) frequency
range ($40-175\text{ GHz}$), we observe two magnonic gaps which already
exist in this system. At around $110\text{ GHz}$, we find a gap resulting
from Bragg scattering of the SWs at the edge of first BZ ($k=\pi/a$) -- marked by the label 'B'. The gap observed at higher frequency ($\sim160\text{ GHz}$), denoted by the label 'AC',
results from the anticrossing of the dispersion branches originating
from the modes of homogeneous waveguide (of the width equal to the
distance between the row of antidots and the edge of the patterned
waveguide \cite{Klo12}). These dispersion branches are plotted in Fig. 2 by black
and gray thick lines. If we artificially double the period to $2a$ then
we obtain the folded branches of dispersion relation (marked by black
dashed and gray dashed lines in Fig. 2a) in narrowed BZ.
At the edge of this BZ ($k=\pi/2a$) the branches intersect
each other. At this point we observe two pairs of degenerate modes.
For the pair with positive (negative) group velocity, the phase increases
with the increase of the distance along the wire -- see corresponding
profile of the modes $1'$ and $2'$ (or $3'$ and $4'$) in Fig. 2b.
For the folded modes (dashed lines) the direction of increase of the
phase is opposite (the profile not shown in Fig. 2b). It is also worth to notice that the SWs in two halves of the waveguide precess olny in-phase (mode
$1'$ or $3'$) or out-of-phase (mode $2'$ or $4'$). 

For non-zero displacement $w$, we can observe (see Fig. 2) the opening
of new magnonic gaps at $k=\pi/2a$. The new gaps, at the frequencies
$\sim85\text{ GHz}$ and $\sim150\text{ GHz}$ become wider with increasing
displacement of the antidots. Note that for higher gap at $\sim150\text{ GHz}$,
a small displacement such as $w=1.5\text{ nm}$ (blue curves) is not
sufficient for opening of gaps, but both the gaps (red bars in Fig.
2) are opened for larger shift of antidots $w=3\text{ nm}$ (red curves).
The opening of the gaps partially lifts the degeneracy between the
two pairs of modes which cross at $k=\pi/2a$ ($w=0$, black and gray
lines). The modes which anticross at the edge of this BZ
($k=\pi/2a$) are still grouped in pairs in the system with SW
pinning. The group velocity of the modes in the range of anticrossing
is significantly reduced (the dispersion branches are practically
flat in this region). This is also manifested in the profiles of SW amplitudes where such modes (see modes $1$, $3$, $5$, $7$
for $w=3\text{ nm}$ in Fig. 2b) have the forms of standing waves
with distinctive zig-zag like nodal lines (white areas) and constant
phase over the localization areas of SWs. 

Figure 3 presents the dispersion relations calculated for the system
with SW pinning at the interfaces between magnetic and nonmagnetic
material calculated by PWM (red dashed) and by MS (grayscale map in
the background). We have selected eight different values of displacement
for the antidots as $w$ = $1.5$, $3$, $4.5$, $6$, $9$, $12$,
$15$ and $18\text{ nm}$. In this range ($1.5\text{ nm}<w<18\text{ nm}$),
the antidots are shifted from the positions closer to the center of
the waveguide ($w=1.5\text{ nm}$) to the locations near the edges
of the waveguide ($w=18\text{ nm}$). For the intermediate values
of the shift $w$, say $\sim6-9\text{ nm}$, the SW has to propagate in meander-like manner. Therefore its scattering is the strongest.  This leads to the increase of the width of the gaps
and to the reduction of the width of the bands. For smaller values
of the shift, where (due to the pinning) the row of antidots almost isolates the SWs
in both halves of the waveguide, the modes appear in almost degenerate
pairs. For larger shifts of antidots, where the SW can propagate
in zig-zag channel between antidots, this degeneracy is significantly
lifted except at the edge of BZ ($k=\pi/2a$). 

The modes at $k=0$ are standing wave modes where the magnetization
precess with spatially uniform phase in the distinctive regions which
are separated by nodal lines  in the spatial profiles
of the modes in Fig. 3b, we can see the spots of uniform colors (representing
the spatially homogeneous phase) which do not join each other with transient colors. The
SW in these regions always precess in-phase or out-of-phase,
with respect to each other. This behavior shows non-propagative characters
of modes at $k=0$ which is also manifested in the dispersion relation
where the dispersion branches become flat at $k=0$. The standing
wave modes (with group velocity equal to zero) are also expected at
the edge of the BZ where patterns of standing waves result
from the interference of counter-propagative waves differing by reciprocal
lattice vector. However, in our system we can find the propagating
SWs exactly at the edge of the BZ. For larger values
of shift of the antidots, the phase changes continuously in wave-like
channels and we do not observe nodal lines for amplitude across the
waveguide. The corresponding dispersion branches are also tilted at the edge of
BZ. Both observations indicate that we should deal with
propagating modes at $k=0.5\pi/a$. It is also worth to notice that
these modes appear at the edge of BZ in doubly degenerate
pairs. This degeneracy can be explained by inspection of the spatial
profiles of dynamical magnetization. It is known that Bloch function
at the edge of BZ flips its phase after translation by
one period. In our case, we have to do with the system with complex
unit cell containing two antidots in one period. These antidots are
placed equidistantly along the waveguide and are shifted by the same
distance toward the opposite edges of the waveguide. Due to this symmetry,
the phases are supposed to change by $+\pi/2$ or $-\pi/2$ during
the translation by each half of period. We observe such behavior in
the profiles of the degenerate modes at the boundaries of BZ where the phase increases (or decreases) along the waveguide.
The calculations presented in Fig. 3 were done for the case of magnetization
pinning. However, similar degeneracy of modes at the boundary of BZ can be observed in SW dispersion for the system where
the magnetization was pinned at the interfaces with nonmagnetic
material \textendash{} see Fig. 4. The modes in each pair are also
counter-propagative at $k~=~0.5\pi/a$. 

\begin{figure}[!ht]
\includegraphics[width=8.5cm]{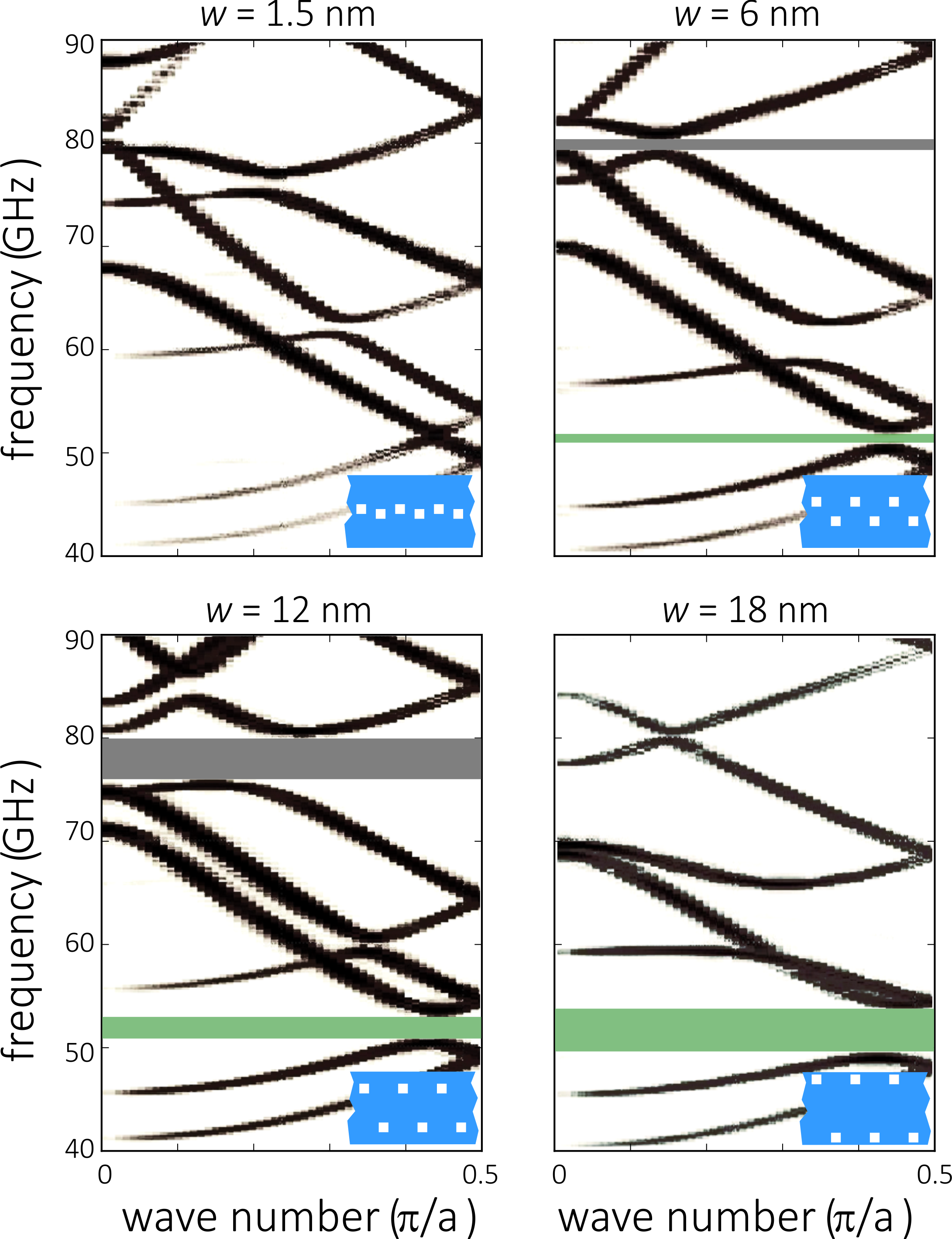}
\caption{Spin wave dispersion relation for different displacements of antidots
with $w=1.5,\text{ }6,\text{ }12,\text{ }18\text{ nm}$ calculated
with the aid of micromagnetic simulations. We assumed that the magnetization
is free at the edges of the waveguide and the antidots. The green and gray
bars denote the lowest magnonic gaps resulting from the presence of
the complex unit cell.}
\end{figure}

In the dispersion relation, we can easily identify the pair of modes which
are degenerate and cross each other with non-zero slope at the edge
of BZ. By inspecting the profiles of SWs, we noticed
that the phase increases in opposite directions along the waveguide
for the each of the modes in every such pair (see e.g., the profiles
of modes 1 and 2 in Fig. 3b for $w$ = 9 and 18). The direction of spatial
changes of phase allows us to classify the modes into two groups.
The dispersion branches of the modes of the different symmetries (manifested
by the opposite directions of the spatial changes of phase) can cross
each other whereas those of the same symmetry have to anti-cross.
For the system with magnetization pinning we can observe (see Fig.
3 for $w=3,\text{ }4.5,\text{ }6\text{ nm}$) that the pairs of modes
which were initially degenerate ($w=0$) in the whole of the BZ
(see Fig. 2) cross each other at few additional points in the BZ. For the system with pinned magnetization
the anti-crossing is clearly visible between second and third mode
(band) for larger values of antidots displacement ($w=18\text{ nm}$)
where the splitting in mentioned pair of modes is significant at $k=0$.
This anti-crossing is responsible for keeping the magnonic gap (marked
by green bar) opened. The effect of crossing (and anti-crossing) of
dispersion branches resulting for the differences (and correspondence)
in symmetry of modes is also observed for system with unpinned magnetization
(see Fig. 4). We can see that, due to the lack of pinning, the SWs are constrained to a lesser extent which results in the weaker
quantization of their modes -- we can notice many more
modes in the same frequency range, referring to the system with
strong pinning (see Fig. 3). Therefore, for the system with considered
sizes where the magnetization is unpinned, the anti-crossing is crucial
to observe the magnonic gap being opened at all. 

The differences between the systems with pinned and unpinned magnetization
are the most striking for small values of displacement of antidots.
For the system with pinning, the waveguide is artificially split into
two half-waveguides by the row of antidots placed in its center. Due
to the weak crosstalk between SW in such half-waveguides,
the SWs eigenmodes are almost degenerate. We can notice that
the lowest dispersion branches appear in overlapping pairs and the
profiles of the modes for each of such pairs show that SWs
in half-waveguides precesses in-phase or out-of-phase (see Fig. 2
and Fig. 3 for $w=1.5\text{ nm}$). The pinning at the antidots edges
at the center of the waveguide enhances the confinement of SWs.
When the antidots are shifted towards the edges of the waveguide,
the zigzag-like channel between them is opened and eventually its
width becomes close to the width of the whole waveguide. Therefore,
with the increase in the displacement of the antidots, the constraints
for SWs become weaker, the modes are quantized denser in frequency
scale and the frequencies of SW modes are shifted gradually
downwards. Strikingly, this effect is absent for the system with the
magnetization released at the interfaces with nonmagnetic material
(see Fig. 4). For this system, the bottom of the lower magnonic band
is located approximately at the same frequency, regardless of the
displacement of the antidots. Moreover, for the case of unpinned magnetization,
the SWs are not separated by the row of antidots even if the
antidots are aligned in the center of the waveguide. Thus, we do not
observe the overlapped pair of dispersion branches at this system
(see Fig. 4).

\begin{figure}[!ht]
\includegraphics[width=8.2cm]{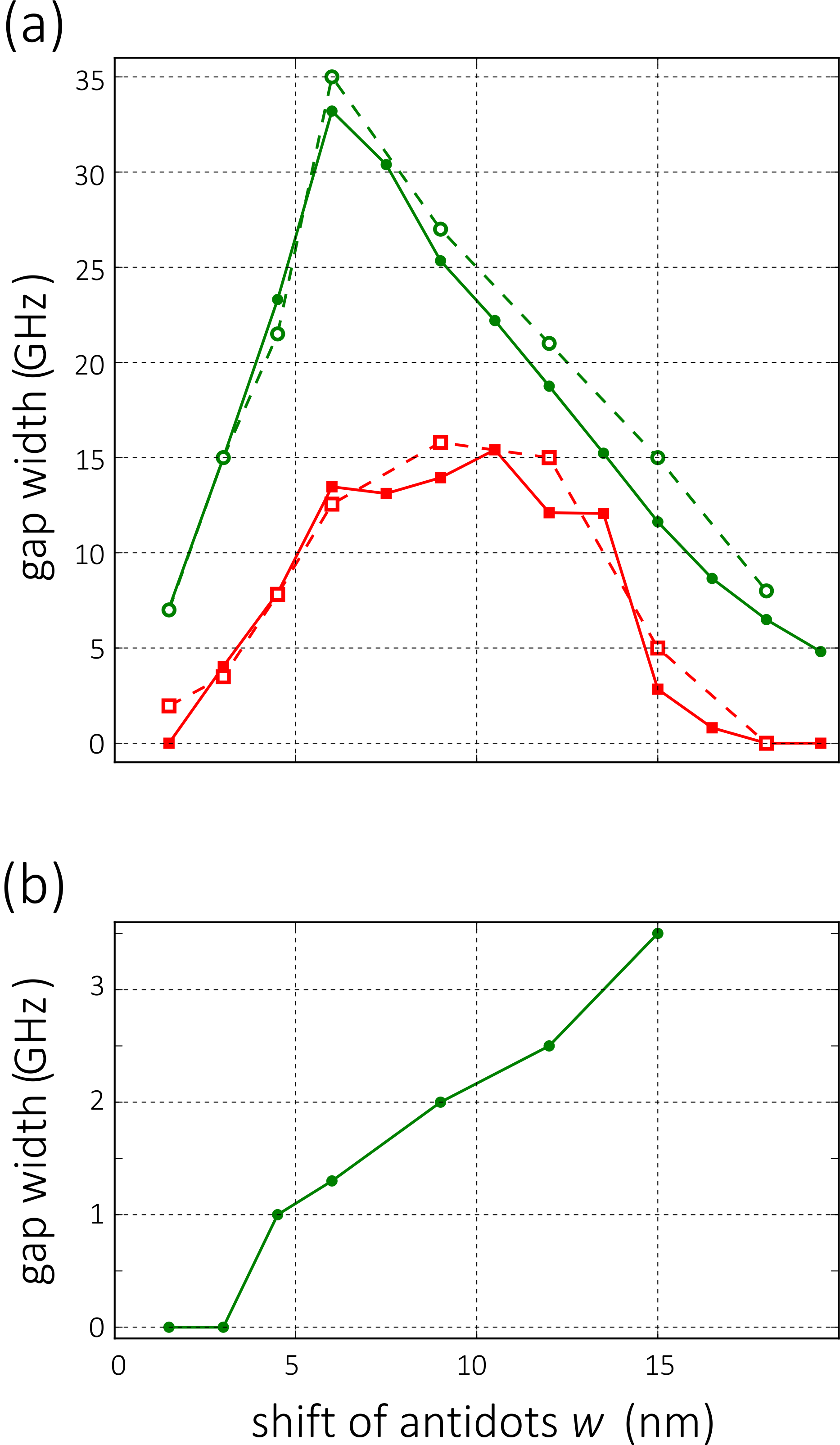}
\caption{Dependence of the width of the first and second magnonic gaps (green and red lines respectively) for the waveguide with pinned magnetization a) and unpinned magnetization b). Dashed and solid lines refer to plane wave method calculations and micromagnetic simulations, respectively.}
\end{figure}

We will discuss now the dependencies of the widths of the magnonic
gaps on the value of the displacement of antidots. We will consider
the gaps resulting from the introduction of complex base of unit cell
only (i.e., the base containing two elements/antidots in unit cell).
Figure 5 presents these dependencies for such gaps marked in Fig. 3
and Fig. 4 by green and pink bars.

The mechanism for the opening of gaps for the system with
pinned magnetization was presented in details in discussion referring
to Fig. 2. We pointed out that new gaps are opened at the edges of
BZ. For the antidots aligned at the center of the waveguide
the BZ was artificially reduced and we got the degeneracy
of modes (crossing of dispersion branches) at the BZ edges resulting
from the folding of dispersion relation. The introduction of displacement
of antidots makes the unit cell containing two antidots essentially
elementary and enhances the coupling between two halves of the waveguide.
This partially lifts the degeneracy of modes, which results in the
opening of new magnonic gaps. Note that at the edge of BZ
the modes remain doubly degenerate, which results from the symmetry
related to the choice of the direction of spatial changes of phase
along the waveguide. 

In the system where the magnetization is unpinned,
the parts of the waveguide on the opposite sides of the sequence of
antidots are strongly coupled regardless on the displacement of the
antidots. Therefore, we observe the degeneracy at BZ edge, 
related only to the choice of two equivalent directions of spatial
changes of phase. The new magnonic gaps are opened due to anti-crossing
of dispersion branches for which the phase increases in the same direction
along the waveguide. The gaps discussed here are induced by presence
of the unit cell of complex base and disappear when this complexity
is artificial, i.e., when there is no displacement of antidots. 
Both the case of pinned and unpinned magnetization the width of discussed
gaps increases for small values of displacement of antidots. However,
this increasing trend is not sustained for system with magnetization
pinning. For this system, the maximum width is reached for displacement
$w\sim6\text{ nm}$ ($w\sim10\text{ nm}$) for the first (second)
magnonic gap \textendash{} see Fig. 5a. It corresponds to the case
when the waveguide is divided by antidots in three sub-waveguides
of comparable width. For the system with pinning, the width of the
gaps decreases when the antidots start to approach the edges of the
waveguide. This behavior can be understood if we notice that, due
to pinning, the amplitude of the SW precession decrease gradually
in the vicinity of the edges of the waveguide. The location of the
antidots in this region will not influence significantly on SW
propagation through the waveguide, and therefore, the SW spectrum
should become similar to the spectrum of uniform waveguide where the
magnonic gaps are not observed. In the waveguide with unpinned magnetization,
the amplitude of SWs reaches the largest values at the edges.
It explains why the scattering of SWs should be significant
and will lead to opening the largest magnonic gaps for the periodic
sequence of antidots placed close to the edges of the waveguide. 

\begin{figure}[!ht]
\includegraphics[width=8.2cm]{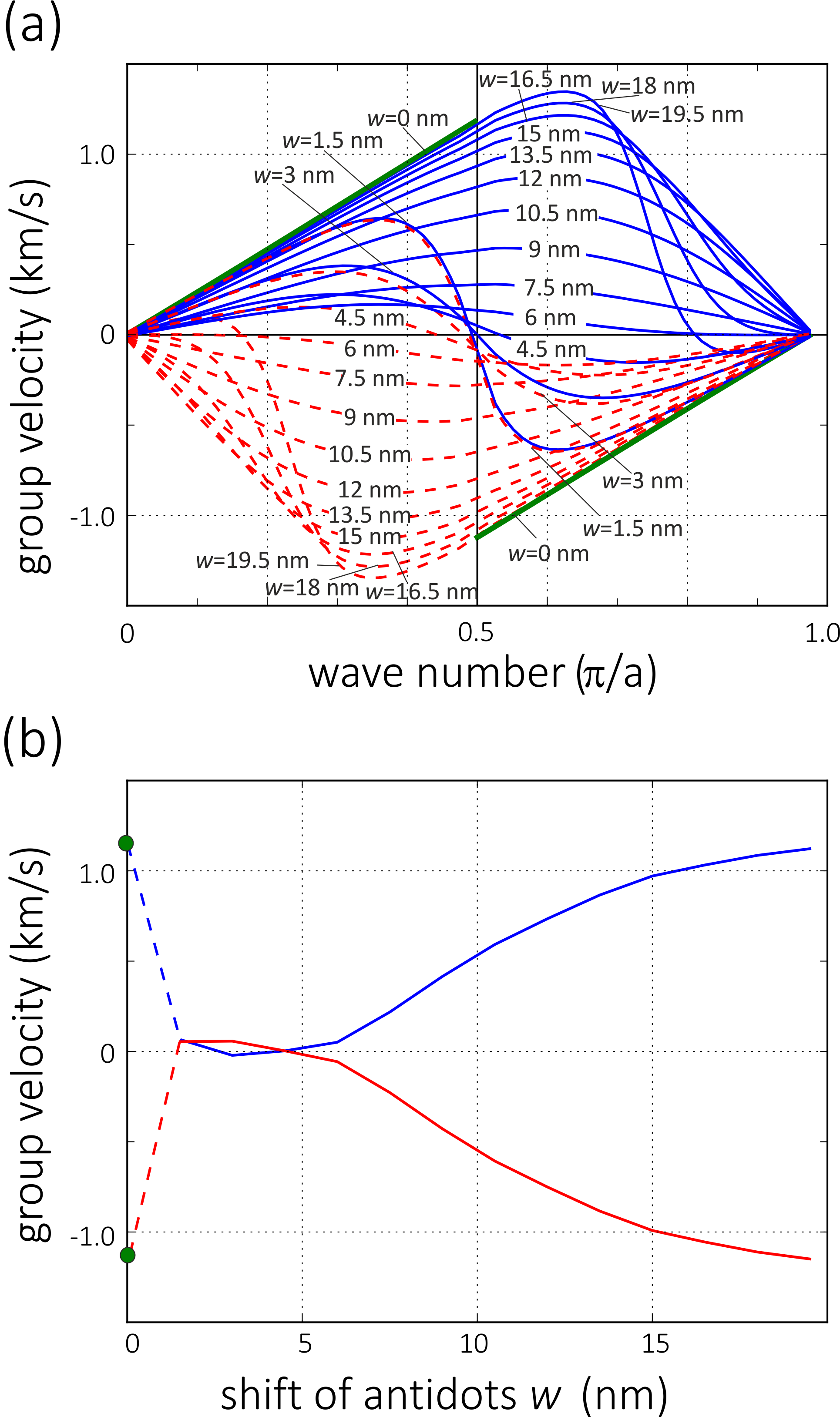}
\caption{
a) Dependence of the group velocity for the first and second band (red dashed and solid blue curves, respectively) on the wave number. The values $k=0$ and $k=0.5\pi/a$ denote the center and the edge of the first Brillouin zone. Dependence of the group velocity for two lowest bands at the edge of the first Brillouin zone as a function of displacement of antidots $w$.}
\end{figure}

Next, we drawn attention to the propagative character of the modes
at the edges of BZ which was manifested both in the SW mode profiles and the non-zero slope of dispersion branches.
To investigate this effect in further details, we numerically calculated
the group velocity for the two lowest dispersion branches of the system
with pinned magnetization. Figure 6a presents the dependence of the
group velocity in this system on the wave number in the first BZ. The thick green lines denote the group velocity for the system
where the antidots are aligned in-line at the center of the waveguide
with artificially extended unit cell containing two antidots (see
Fig. 2 for reference). We can see that the group velocity changes
practically linearly from the center to the edge of the BZ, which proves the quadratic dispersion relation, that is a characteristic
for the exchange dominated regime. For tiny displacements of antidots
the gap is just opened and we observe maximum of dispersion relation (dependence of the frequency on wave number: $f(k)$) 
at the edge of the BZ. As a result, the group velocity $v_{g}=2\pi\frac{d f}{d k}$
drops to zero at $k~=~0.5\pi/a$ but for smaller values of wave number
$k$, the group velocity changes  linearly with $k$. The increase
of antidots displacement (up to $\sim 6\text{ nm}$) makes the scattering
of SWs stronger, which increases the width of the gap, reduces
the width of the bands and decreases the group velocity. Above this
limit of displacement, the SWs of lowest modes start to propagate
through the zigzag-like channel between displaced antidots. This causes
three effects: (i) we do not observe degeneracy of SW modes
expected at the edge of BZ -- this degeneracy was related
to the separation of modes between two half-waveguides by the row
of antidots, (ii) we can notice that two lowest SW modes of
the same wave number can propagate in opposite directions \textendash{}
i.e., they have opposite group velocities (this counter-propagative
modes are marked in Fig. 6a by red and blue-dashed lines) and (iii)
the absolute value of group velocity increases with the further increase
of the antidots displacement \textendash{} see Fig. 6b. 

\section{Conclusions}

In summary, we have systematically investigated the magnonic band
structure in a planar magnonic waveguide with periodic modulation
of the antidot position across the width of the waveguide. Our study
reveals the possibilities to open-up new magnonic band gap and control
their position and width in frequency and wavevector domains by the introduction
of two antidots in one unit cell (i.e. for the unit cell with complex base).
We showed that the  folding of the BZ by introducing
double periodicity results in anti-crossing of the modes of different symmetry
and opening of new magnonic gap. The width of the gap can be controlled
by varying the shift of the position of the antidot from the long
axis of the waveguide. We investigated the antidot position modulated
gap width for both strong SW pinning and ideally unpinned magnetization
at the material/air interface. We found a non-monotonic dependence
of the gap width on the shift of the position of the antidot in the
first case, whereas the gap width increases in a regular manner for
the latter case. This is understood in terms of strong (or weak) SW scattering for the antidots shifted close to the edges of the waveguide without (or with pinning). Moreover, we showed that anti-crossing is crucial for opening of magnonic
band gap in case of unpinned magnetization where the Bragg scattering
is not so strong. Interestingly, we observed a continuous change in
phase for the lowest frequency modes at the edge of the BZ as an indication of propagative character. Further in depth investigation
reveals that the lowest modes in folded BZ have significantly large
group velocity at the edge of the BZ. Strikingly, the group velocity
has opposite sign in each degenerated pair of modes at the edge of BZ and can be molded by the mentioned shift of the antidots. Our findings unveil a new way to design
the magnonic waveguide with suitable complex base for SW propagation.
By designing this kind of nanoscale waveguide structure, we can create
and annihilate the magnonic band gap at very high frequency, which
is key for the spectral filter application, as well as control the
propagation velocity and phase change of SWs, essential for
the design of phase shifter and delay generator.
 
\begin{acknowledgments}
This work was supported by Indo-Polish joint Project DST/INT/POL/P-11/2014,
the National Science Centre Poland grant UMO-2012/07/E/ST3/00538,
and the EU\textquoteright s Horizon2020 research and innovation programme
under the Marie Sklodowska-Curie GA No. 644348(MagIC). S.P. acknowledges
Department of Science and Technology, Govt. of India for INSPIRE fellowship.

\end{acknowledgments}

\bibliography{bibliography}

\end{document}